\documentclass[12pt]{article}
\usepackage{amsmath,array,hyperref}
\numberwithin{equation}{section}

\renewcommand{\cases}[1]{\left\{\begin{array}{ll} #1 \end{array}\right.}
\renewcommand{\title}[1]{{\Large\bf\mbox{}\\\medskip#1\bigskip\medskip\\ }}
\renewcommand{\author}[1]{{\large #1\smallskip\\ }}
\newcommand{\address}[1]{{\em #1\medskip\\ }}
\newcommand{\comment}[1]{}

\def\swnedots{\mathinner{\mkern1mu\raise1pt\vbox{\kern7pt\hbox{.}}\mkern2mu
\raise4pt\hbox{.}\mkern2mu\raise7pt\hbox{.}\mkern1mu}}
\def\slhat{\widehat{s\ell}}

\def\({\left(}
\def\){\right)}
\newcommand{\ds}{\displaystyle}

\newcommand{\mi}{\!-\!}
\newcommand{\plus}{\!+\!}

\font\tenmsb=msbm10 scaled \magstep 2
\font\sevenmsb=msbm10 scaled \magstep 1
\font\fivemsb=msbm10
\newfam\msbfam
\textfont\msbfam=\sevenmsb
\scriptfont\msbfam=\sevenmsb
\scriptscriptfont\msbfam=\fivemsb
\def\Bbb#1{{\fam\msbfam\relax#1}}
\newfam\smmsbfam
\textfont\smmsbfam=\sevenmsb

\newfam\Bmsbfam
\textfont\Bmsbfam=\tenmsb

\font\bigtenmsb=msbm10 scaled \magstep 3
\font\bigsevenmsb=msbm7 scaled \magstep 3
\font\bigfivemsb=msbm5 scaled \magstep 3
\newfam\bigmsbfam
\textfont\bigmsbfam=\bigtenmsb
\scriptfont\bigmsbfam=\bigsevenmsb
\scriptscriptfont\bigmsbfam=\bigfivemsb

\newcommand{\W}[2]{\!\left(\ds\left.\!\begin{array}{ccc}#1\end{array}
\right #2\right)}
\newcommand{\Wt}[1]{\!\left(\ds\!\begin{array}{ccc}#1\end{array}\right)}
\newcommand{\smallW}[2]{\!\bigl(\newcolumntype{t}{>{\!}c<{\!}}
\begin{array}{ttt}#1\end{array}\!\bigm #2\bigr)}

\renewcommand{\ss}{\scriptstyle}
\newcommand{\sss}{\scriptscriptstyle}

\newcommand{\p}[2]{\makebox(0,0)[#1]{$#2$}}
\newcommand{\pp}[2]{\makebox(0,0)[#1]{$\ss#2$}}

\renewcommand{\vec}[1]{\mbox{\boldmath$#1$}}

\begin{document}
\begin{center}
\title{Integrable Lattice Realizations \\ of Conformal  Twisted
Boundary Conditions}
\author{C.H. Otto Chui, Christian Mercat, William P.
Orrick and Paul A. Pearce}
\address{Department of Mathematics and Statistics\\
University of Melbourne\\Parkville, Victoria 3010, Australia}
\medskip
\begin{abstract}
\noindent We construct integrable lattice realizations of conformal
twisted boundary conditions for $\slhat(2)$ unitary minimal models on
a torus.  These conformal field theories are realized as the continuum
scaling limit of critical $A$-$D$-$E$ lattice models with positive
spectral parameter.  The integrable seam boundary conditions are
labelled by $(r,s,\zeta)\in(A_{g-2},A_{g-1},\Gamma)$ where $\Gamma$ is
the group of automorphisms of the graph $G$ and $g$ is the Coxeter
number of $G=A,D,E$.  Taking symmetries into account, these are
identified with conformal twisted boundary conditions of Petkova and
Zuber labelled by $(a,b,\gamma)\in(A_{g-2}\otimes G,A_{g-2}\otimes
G,{\Bbb Z}_{2})$ and associated with nodes of the minimal analog of
the Ocneanu quantum graph.  Our results are illustrated using the
Ising $(A_2,A_3)$ and 3-state Potts $(A_4,D_4)$ models.
\end{abstract}
\end{center}

\section{Introduction}

There has been much recent
progress~\cite{BePZ,BPPZ98,BPPZ00,BP01,PZ00,Coq00,PZ0011021,
MP,PZ01,PR01
}
on understanding integrable boundaries in statistical mechanics,
conformal boundary conditions in rational conformal field theories and
the intimate relations between them on both the cylinder and the
torus.  Indeed it appears that, for certain classes of theories, all
of the conformal boundary conditions on a cylinder can be realized as
the continuum scaling limit of integrable boundary conditions for the
associated integrable lattice models.  For $\slhat(2)$ minimal
theories, a complete classification has been
given~\cite{BePZ,BPPZ98,BPPZ00} of the conformal boundary conditions
on a cylinder.  These are labelled by nodes $(r,a)$ of a tensor
product graph $A\otimes G$ where the pair of graphs $(A,G)$, with $G$
of $A$-$D$-$E$ type, coincide precisely with the pairs in the
$A$-$D$-$E$ classification of Cappelli, Itzykson and Zuber~\cite{CIZ}. 
Moreover, the physical content of the boundary conditions on the
cylinder has been ascertained~\cite{BP01,MP} by studying the related
integrable boundary conditions of the associated $A$-$D$-$E$ lattice
models~\cite{Pas} for both positive and negative spectral parameters,
corresponding to \emph{unitary minimal theories} and
\emph{parafermionic theories} respectively.  Recently, the lattice
realization of integrable and conformal boundary conditions for $N=1$
\emph{superconformal theories}, which correspond to the \emph{fused}
$A$-$D$-$E$ lattice models with positive spectral parameter, has also
been understood in the diagonal case~\cite{PR01}.

In this letter, we use fusion to construct integrable realizations of
conformal twisted boundary conditions on the torus~\cite{Coq00,PZ0011021}. 
Although the methods are very general we consider $\slhat(2)$ unitary
minimal models for concreteness.  The key idea is that fused blocks of
elementary face weights on the lattice play the role of the local
operators in the theory.  The integrable and conformal boundary
conditions on the cylinder are constructed~\cite{BP01} by acting
with these fused blocks on the simple integrable boundary condition
representing the vacuum.  By the generalized Yang-Baxter equations,
these fused blocks or seams can be propagated into the bulk without
changing the spectrum of the theory.  The seams so constructed provide
integrable and conformal boundary conditions on the torus.  One
subtlety with this approach is that fixed boundary conditions $a\in G$
on the edge of the cylinder are propagated into the bulk by the action
$a=\zeta(1)$ of a graph automorphism $\zeta$, which preserves the
Yang-Baxter structure, on the distinguished (vacuum) node $1\in G$. 
In general, for rational conformal field theories on the torus, we
expect the fusions supplemented by the automorphisms to generate all
of the integrable and conformal seams.  We illustrate our approach in
this letter by using the Ising $(A_2,A_3)$ and 3-state Potts
$(A_4,D_4)$ models as examples.  A detailed consideration of the
$A$-$D$-$E$ unitary minimal models will be given in a forthcoming
paper~\cite{CMOP2}.

\section{Lattice Realization of Twisted Boundary Conditions}
\label{sec:Lattice}

\subsection{$A$-$D$-$E$ lattice models and integrable seam weights}
\label{sec:ADE}

The $\slhat(2)$ unitary minimal theories~\cite{BPZ} are realized as
the continuum scaling limit of critical $A$-$D$-$E$ lattice
models~\cite{Pas}.  The spin states $a,b,c,d$ are nodes of a graph
$G=A,D,E$ with Coxeter number $g$.  The bulk face weights are
\begin{equation}
\setlength{\unitlength}{8mm}
W\W{d&c\\ a&b}{|u}\;=\;
\begin{picture}(2,0)(0,.85)\multiput(0.5,0.5)(1,0){2}{\line(0,1){1}}
\multiput(0.5,0.5)(0,1){2}{\line(1,0){1}}
\put(0.48,0.63){\pp{bl}{\searrow}}
\put(0.45,0.45){\pp{tr}{a}}\put(1.55,0.45){\pp{tl}{b}}
\put(1.55,1.55){\pp{bl}{c}}\put(0.45,1.55){\pp{br}{d}}
\put(1,1){\pp{}{u}}\put(2.1,0.8){\p{}{}}\end{picture}
\;=
\;s_{1}(-u)\:\delta_{ac}\;+\;
s_{0}(u)
\;\frac{\sqrt{\psi_a\:\psi_c}}{\psi_b}\:\delta_{bd}
\end{equation}
where $u$ is the spectral parameter with $0<u<\lambda$,
$\lambda=\pi/g$ is the crossing parameter,
$s_{k}(u)={\sin(u+k\lambda)}/{\sin \lambda}$ and the weights vanish if
the adjacency condition is not satisfied on any edge.  The crossing
factors $\psi_a$ are the entries of the Perron-Frobenius eigenvector
of the adjacency matrix $G$.  Note that if $\lambda-\pi/2<u<0$, the
continuum scaling limit describes the principal ${\Bbb Z}_k$
parafermions with $k=g-2$~\cite{MP}.

The $A$-$D$-$E$ lattice models are Yang-Baxter integrable~\cite{Bax}
on a cylinder in the presence of a boundary~\cite{BPO'B96} with
boundary conditions labelled~\cite{BPPZ00} by $(r,a)\in (A_{g-2},G)$. 
A general expression for the $(r,a)$ boundary weights is given
in~\cite{BP01}.  They are constructed by starting at the edge of
the lattice with a fixed node $a\in G$ and adding a fused block of
$r-1$ columns.  Strictly speaking, this construction on the cylinder
is implemented with double row transfer matrices.  Nevertheless, our
idea is to propagate these boundary conditions into the bulk using a
description in terms of single row transfer matrices.  The
expectation, which we confirm numerically, is that these integrable
boundary conditions continue to be ``conformal" in the bulk.

For $(r,s,\zeta)\in(A_{g-2},A_{g-1},\Gamma)$, we define integrable seam
weights ${W^{(r,s,\zeta)}}\smallW{d&\gamma&c\\ a&\alpha&b}{|\!u, \xi}$
by taking the fusion product of three seams of types
$(r,s,\zeta)=(r,1,1)$, $(1,s,1)$, $(1,1,\zeta)$ respectively.  The
$(r,1,1)$ seam is obtained by fusing $r-1$ columns or faces
\setlength{\unitlength}{10mm}
\begin{equation}
\begin{picture}(8,3)
      \put(0,1.5){\pp{c}{{\ds W^{(r,1,1)}}\W{d&\gamma&c\\ 
a&\alpha&b}{|u,\xi}\;\;=\;\;
      {\ds\prod_{k=1}^{r-2}
      s_{-k}(u+\xi)^{-1}}
      }}
      \put(3.5,0.5){
\begin{picture}(5,2)
\multiput(0.5,0.5)(0,1){2}{\line(1,0){4}}
\multiput(0.5,0.5)(1,0){2}{\line(0,1){1}}
\multiput(3.5,0.5)(1,0){2}{\line(0,1){1}}
\multiput(0.48,0.58)(3,0){2}{\pp{bl}{\searrow}}
\put(4,1){\pp{c}{u+\xi}}
\put(2.5,1){\pp{c}{\cdots}}
\put(0.6,1.17){\pp{l}{u+\xi-}}\put(1.45,0.9){\pp{r}{(r-2)\lambda}}
\put(0.48,0.48){\pp{tr}{a}}\put(4.52,0.48){\pp{tl}{b}}
\put(0.48,1.52){\pp{br}{d}}\put(4.52,1.52){\pp{bl}{c}}
\put(1.5,0.39){\pp{t}{e_1}}\put(3.6,0.39){\pp{t}{e_{r-2}}}
\put(1.5,1.64){\pp{b}{g_1}}\put(3.6,1.64){\pp{b}{g_{r-2}}}
\multiput(1.5,0.5)(2,0){2}{\pp{}{\bullet}}
\multiput(1.5,1.5)(2,0){2}{\pp{}{\bullet}}
\end{picture}}
\put(6,0){\p{b}{U^r(a,b)_{\alpha,(a,e_1,\ldots,e_{r\mi2},b)}}}
\put(6,3){\p{t}{U^r(d,c)_{\gamma,(d,g_1,\ldots,g_{r\mi2},c)}}}
\end{picture}
      \label{eq:Wr1}
\end{equation}
These weights depend on the external spins $a,b,c,d\in G$ and on the
internal {\em bond variables} $\alpha,\gamma$ labelling the fused
edges~\cite{ZP94}.  The remaining internal spins indicated with solid
dots are summed out.  Here $\alpha= 1,2,\ldots,F^{r}_{a\,b}$ and
$\gamma=1,2,\ldots,F^{r}_{c\,d}$ where the fused adjacency matrices
$F^{r}$ with $r=1,2,\ldots,g-2$ are defined recursively in terms of
the adjacency matrix $G$ by the $s\ell(2)$ fusion rules
\begin{eqnarray}
F^{1}=I,\qquad F^2=G,\qquad F^r=GF^{r-1}-F^{r-2}
\label{eq:F}
\end{eqnarray}
where the superscript is an index and not a matrix power.  The fused
adjacency matrices $F^r$ are related to the usual intertwiners $V_r$
and operator content $n_r$ on the cylinder by
$F^r_{a\,b}=V_{ra}{}^b=n_{ra}{}^b$.  The seam weights vanish if
$F_{ab}^r F_{bc}^2 F_{cd}^r F_{da}^2=0$ and the fusion is implemented
via the fusion vectors $U^r$ listed in \cite{BP01}.  The
inhomogeneity or seam field $\xi$ is arbitrary and can be taken
complex but $\xi$ must be specialized appropriately for the seam
boundary condition to be conformal.  Although we usually take
$\xi=-3\lambda/2$ at the isotropic point $u=\lambda/2$, in fact we
obtain the same twisted partition function for $\xi$ in a suitable
real interval.

The $(1,s,1)$ seam weights are independent of $u$ and $\xi$ and are
given by the \emph{braid limit} $\xi\to i\infty$ of the $(r,1,1)$ seam
weights divided by $i^{s-1} s_{0}(u+\xi)$
\begin{equation}
W^{(1,s,1)}\W{d&\gamma&c\\ a&\alpha&b}{.}\;=\;
(-ie^{-i\frac{\lambda}{2}})^{s-1}\lim_{\xi\rightarrow i\infty}
{s_0(u+\xi)}^{-1}\,
\ds W^{(s,1,1)}\W{d&\gamma&c\\ a&\alpha&b}{|u,\xi}
\end{equation}
The automorphisms $\zeta\in\Gamma$ of the adjacency matrix, satisfying
$G_{a,b}=G_{\zeta(a),\zeta(b)}$, leave the face weights invariant
\begin{equation}
\setlength{\unitlength}{8mm}
W\W{d&c\\ a&b}{|u}\;=\;
\begin{picture}(2,0)(0,.85)\multiput(0.5,0.5)(1,0){2}{\line(0,1){1}}
\multiput(0.5,0.5)(0,1){2}{\line(1,0){1}}
\put(.5,.52){\pp{bl}{\sss \searrow}}
\put(0.45,0.45){\pp{tr}{a}}\put(1.55,0.45){\pp{tl}{b}}
\put(1.55,1.55){\pp{bl}{c}}\put(0.45,1.55){\pp{br}{d}}
\put(1,1){\pp{}{u}}\put(2.1,0.8){\p{}{}}\end{picture}
\;=\;
\begin{picture}(2,0)(0,.85)\multiput(0.5,0.5)(1,0){2}{\line(0,1){1}}
\multiput(0.5,0.5)(0,1){2}{\line(1,0){1}}
\put(.5,.52){\pp{bl}{\sss \searrow}}
\put(0.45,0.45){\pp{tr}{\zeta(a)}}\put(1.55,0.45){\pp{tl}{\zeta(b)}}
\put(1.55,1.55){\pp{bl}{\zeta(c)}}\put(0.45,1.55){\pp{br}{\zeta(d)}}
\put(1,1){\pp{}{u}}\put(2.1,0.8){\p{}{}}\end{picture}
\;=\;W\W{\zeta(d)&\zeta(c)\\ \zeta(a)&\zeta(b)}{|u}
\end{equation}
and act through the special seam~\cite{CKP}
\begin{equation}
\setlength{\unitlength}{8mm}
      {\ds W^{(1,1,\zeta)}}\Wt{d&c\\ a&b}\;=
\begin{picture}(2,0)(0,.85)\multiput(0.5,0.5)(1,0){2}{\line(0,1){1}}
\multiput(0.5,0.5)(0,1){2}{\line(1,0){1}}
\put(.5,.52){\pp{bl}{\sss \searrow}}
\put(0.45,0.45){\pp{tr}{a}}\put(1.55,0.45){\pp{tl}{\zeta(a)}}
\put(1,1){\pp{}{\zeta}}
\put(1.55,1.55){\pp{bl}{\zeta(d)}}\put(0.45,1.55){\pp{br}{d}}
\put(1,1){\pp{}{}}\put(2.1,0.8){\p{}{}}\end{picture}
\;=\;
\cases{1,& b=\zeta(a),\ c=\zeta(d)\cr 0,& \mbox{otherwise.}}
      \label{eq:seamWzeta}
\end{equation}
Notice that the $(r,s,\zeta)=(1,1,1)$ seam, where $\zeta=1$ denotes
the identity automorphism, is the empty seam corresponding to periodic
boundary conditions
\begin{equation}
   W^{(1,1,1)}\W{d&c\\ a&b}{.}\;=
\;\delta_{ab}\delta_{cd}F_{b\,c}^2.
      \label{eq:W11}
\end{equation}

The $A$-$D$-$E$ face and seam weights satisfy the generalized
Yang-Baxter and boundary Yang-Baxter equations ensuring commuting row
transfer matrices and integrability with an arbitrary number of seams. 
Also by the generalized Yang-Baxter equation, the $(r,1,1)$ and
$(1,s,1)$ seams can be propagated along a row and even pushed through
one another without changing the spectrum.  The conformal partition
functions with multiple seams are described by the fusion algebra. 
Explicit expressions can be obtained for the $(r,s,\zeta)$ seam
weights.  The $(2,1,1)$-seam corresponds to a single column with
spectral parameter $u+\xi$ and the $(1,2,1)$ seam, given by the braid
limit, has weights
\begin{equation}
W^{(1,2,1)}\W{d&c\\ a&b}{.}\;=
\;i\,e^{i{\lambda\over 2}}\:\delta_{ac}\;-\;
i\,e^{-i{\lambda\over 2}}
\;\frac{\sqrt{\psi_a\:\psi_c}}{\psi_b}\:\delta_{bd}\,,
      \label{eq:W12}
\end{equation}
More generally, we find that
\begin{eqnarray}
      {\ds W^{(r,1,1)}}\W{d&\gamma&c\\ a&\alpha&b}{|u,\xi}&
      \!\!\!=\!\!\!&
     S_{r-1}s_{1}(u+\xi)\, U^{r}\smallW{d&\gamma&c\\ a&\alpha&b}{.} -
     S_{r}s_{0}(u+\xi)\, V^{r}\smallW{d&\gamma&c\\ a&\alpha&b}{.}\\
      {\ds W^{(1,s,1)}}\W{d&\gamma&c\\ 
a&\alpha&b}{|u,\xi}&\!\!\!=\!\!\!&i^{s-1}\left[
     S_{s-1}\,e^{i{\lambda\over 2}}\, 
     U^{s}\smallW{d&\gamma&c\\ a&\alpha&b}{.} -
     S_{s}\,e^{-i{\lambda\over 2}}\,
     V^{s}\smallW{d&\gamma&c\\ 
a&\alpha&b}{.}\right]\qquad
     \label{eq:WsUV}
\end{eqnarray}
where $S_{k}=s_{k}(0)$ and, in terms of the fusion vectors $U^r$ 
listed in \cite{BP01},
the elementary fusion faces are
\begin{eqnarray}
     U^{r}\smallW{d&\gamma&c\\ a&\alpha&b}{.}  & = &
     \sum_{(d,a,e_{1},\ldots,e_{r\mi3},c,b)}
     U^{r}_{\gamma}(d,c)_{\sss (d,a,e_{1},\ldots,e_{r\mi3},c)}
     U^{r}_{\alpha}(a,b)_{\sss (a,e_{1},\ldots,e_{r\mi3},c,b)}
      \label{eq:Ur}  \\
      V^{r}\smallW{d&\gamma&c\\ a&\alpha&b}{.} & = &
     \sum_{(d,a,e_{1},\ldots,e_{r\mi3},c,b)}
     U^{r}_{\gamma}(d,c)_{\sss (d,a,e_{1},\ldots,e_{r\mi3},c)}
     U^{r}_{\alpha}(a,b)_{\sss
     (a,e_{1},\ldots,e_{r\mi3},c,b)}\times
     \label{eq:Vr}\\
     && \qquad \qquad \qquad \qquad
     \sum_{\beta=1}^{F^{r\plus 1}_{bd}}
     U^{r\plus 1}_{\beta}(d,b)_{\sss
     (d,a,e_{1},\ldots,e_{r\mi3},c,b)}
     \notag
\end{eqnarray}

\subsection{Transfer matrices and symmetries}\label{sec:transfer}

The entries of the transfer matrix with an $(r,s,\zeta)$ seam are
given by
\begin{gather}
      \mbox{}\hspace{-.5in}
\rule{0pt}{24pt}\langle \vec{a},\alpha_{r},\alpha_{s}|\;
\vec{T}_{(r,s,\zeta)}(u,\xi)\;
       |\vec{b},\beta_{r},\beta_{s}\rangle=\;
\setlength{\unitlength}{10mm}
\begin{picture}(7,0)(0,.85)
     \multiput(0.5,0.5)(0,1){2}{\line(1,0){7}}
\multiput(0.5,0.5)(1,0){2}{\line(0,1){1}}
\multiput(3.5,0.5)(1,0){5}{\line(0,1){1}}
\put(0.48,0.63){\pp{bl}{\searrow}}
\multiput(3,0)(1,0){4}{\put(0.48,0.63){\pp{bl}{\searrow}}}
\put(0.5,0.35){\pp{c}{a_{1}}}
\put(1.5,0.35){\pp{c}{a_{2}}}
\put(3.5,0.35){\pp{c}{a_{N}}}
\put(4.5,0.35){\pp{c}{a_{N\plus1}}}
\put(5,0.5){\pp{c}{\alpha_{r}}}
\put(5.5,0.35){\pp{c}{a_{N\plus2}}}
\put(6,0.5){\pp{c}{\alpha_{s}}}
\put(6.5,0.35){\pp{c}{a_{N\plus3}}}
\put(7.5,0.35){\pp{c}{a_{1}}}
\put(0.5,1.7){\pp{c}{b_{1}}}
\put(1.5,1.7){\pp{c}{b_{2}}}
\put(3.5,1.7){\pp{c}{b_{N}}}
\put(4.5,1.7){\pp{c}{b_{N\plus1}}}
\put(5,1.5){\pp{c}{\beta_{r}}}
\put(5.5,1.7){\pp{c}{b_{N\plus2}}}
\put(6,1.5){\pp{c}{\beta_{s}}}
\put(6.5,1.7){\pp{c}{b_{N\plus3}}}
\put(7.5,1.7){\pp{c}{b_{1}}}
\put(1,1){\pp{}{u}}
\put(2.5,1){\pp{}{\cdots}}
\put(4,1){\pp{}{u}}
\put(5,1.2){\pp{}{W^{r,1,1}}}\put(5,.8){\pp{}{(u,\xi)}}
\put(6,1){\pp{}{W^{1,s,1}}}
\put(7,1){\pp{}{W^{1,1,\zeta}}}
\put(2.1,0.8){\p{}{}}\end{picture}
       \label{eq:transfer} \notag \\[16pt]
     =\;\prod_{j=1}^{N}
W\W{b_{j}&b_{j+1}\\ a_{j}&a_{j+1}}{|u}\times\\
W^{(r,1,1)}\W{b_{N+1}&\beta_{r}&b_{N+2}\\
       a_{N+1}&\alpha_{r}&a_{N+2}}{|u,\xi}
      W^{(1,s,1)}\W{b_{N+2}&\beta_{s}&b_{N+3}\\
      a_{N+2}&\alpha_{s}&a_{N+3}}{.}
W^{(1,1,\zeta)}\Wt{b_{N+3}&b_{1}\\
      a_{N+3}&a_{1}}
    \notag
\end{gather}
By the generalized Yang-Baxter equations these form a one-parameter
family of commuting transfer matrices.  The transfer matrices with a
seam are not translationally invariant, however using the generalized
Yang-Baxter equations and a similarity transformation, the seam can be
propagated along the row leaving the spectrum invariant.  This is the
analog of the property that the twisted partition functions are
invariant under deformation of the inserted defect lines.

The seam weights and transfer matrices satisfy certain symmetries as a
consequence of the usual crossing and reflection symmetries of the
face weights.  For the $(1,s,1)$ seam we find the crossing symmetry
\begin{equation}
W^{(1,s,1)}\W{d&\gamma&c\\ a&\alpha&b}{.}\;=\;
\sqrt{\frac{\psi_a\psi_c}{\psi_b\psi_d}}\,
W^{(1,s,1)}\W{a&\alpha&b\\ d&\gamma&c}{.}^\ast
\end{equation}
so that for real $u$
\begin{equation}
\vec{T}^{(1,s,1)}(\lambda-u)\;=\;{\vec{T}^{(1,s,1)}}(u)^\dagger
\end{equation}
and the $(1,s,1)$ transfer matrices are normal matrices.  In
particular, at the isotropic point $u=\lambda/2$ the $(1,s,1)$
transfer matrices are hermitian and the eigenvalues are real.  In
contrast, for the $(r,1,1)$ seam, $\vec{T}^{(r,1,1)}(u,\xi)$ is not
normal in general due to the parameter $\xi$.  However, at
$\xi=\xi_k=\frac{\lambda}{2}(r-2+kg)$ with $k$ even we find
\begin{equation}
\vec{T}^{(r,1,1)}(\lambda-u,\xi_k)\;=\;{\vec{T}^{(r,1,1)}}(u,\xi_k)^T
\end{equation}
so for $\xi=\xi_k$ the transfer matrices are normal.  In this case the
transfer matrices are real symmetric at the isotropic point
$u=\lambda/2$ and the eigenvalues are again real.

\subsection{Finite-size corrections}
\label{sec:FiniteSize}
In the scaling limit the $A$-$D$-$E$ lattice models reproduce the
conformal data of the unitary minimal models through finite-size
corrections to the eigenvalues of the transfer matrices.  If we fix
$\xi$ to a conformal value and write the eigenvalues of the row
transfer matrix $\vec{T}^{(r,s,\zeta)}(u,\xi)$ as
\begin{equation}
T_n(u)=\exp(-E_n),\quad n=0,1,2,\ldots
\end{equation}
then to order $o(1/N)$ the finite-size corrections to the energies
$E_n$ take the form
\begin{eqnarray}
E_0&=&Nf(u)+f_{r,s}(u)-{\pi c\over 6N}\sin\vartheta\\
E_n-E_0&=&
\frac{2\pi}{N}\,\left[(\Delta_n+\bar\Delta_n+k_n+\bar{k}_n)\sin\vartheta
+i(\Delta_n-\bar\Delta_n+k_n-\bar{k}_n)\cos\vartheta\right]
\end{eqnarray}
where $f(u)$ is the bulk free energy, $f_{r,s}(u,\xi)$ is the seam free
energy, $c$ is the central charge, $\Delta_n,\bar\Delta_n$ are
conformal weights, $\vartheta=g u$ is the anisotropy angle and
$k_n,\bar{k}_n\in{\Bbb N}$.  The bulk and seam free energies depend on
$G$ only through the Coxeter number $g$ and are independent of $s$ and
$\zeta$.  To determine $f(u)$ and $f_r(u)$ it is therefore sufficient
to consider $A$-type graphs.

For $G=A_L$, the transfer matrices
$\vec{T}(u)=\vec{T}^{(r,s,\zeta)}(u,\xi)$ satisfy universal TBA
functional equations~\cite{CMP} independent of the boundary conditions
$(r,s,\zeta)$.  It follows that the eigenvalues satisfy the inversion
relation
\begin{equation}
          T(u) T(u+\lambda)=(-1)^{s-1}
s_{1}(u+\xi)s_{1-r}(u+\xi)\left[s_{1}(u) s_{-1}(u) \right]^{N}
      \label{eq:InversionRelation}
\end{equation}
Using appropriate analyticity properties, it is straightforward to
solve~\cite{Stro,Bax82,O'BP97} this relation to obtain closed 
formulas for the bulk and seam
free energies but we do not give the formulas here. Removing the bulk 
and boundary seam contributions to the partition function
on a torus allows us to obtain numerically the conformal twisted 
partion functions
$Z_{(r,s,\zeta)}(q)$ for an $M\times N$ lattice where
$q=\exp(2\pi i\tau)$ is the modular parameter and 
$\tau=(M/N)\exp[i(\pi-\vartheta)]$.

\section{Conformal Twisted Partition Functions}\label{sec:conf}

The conformal twisted partition functions for a rational CFT with a
seam $x=(a,b,\gamma)$ have been given by Petkova and
Zuber~\cite{PZ0011021} as
\begin{equation}
Z_x(q)=\sum_{i,j} {\tilde V}_{ij^*;1}{}^x
\chi_{i}(q){\chi_j(q)}^*,\qquad\qquad
{\tilde V}_{ij^*;1}^{(G)}{}^x=\sum_{c\in T_\gamma} n_{ic}{}^a n_{jc}{}^b
\end{equation}
where $a,b\in G$ and $T_\gamma$, possibly depending on a ${\Bbb Z}_2$
automorphism $\gamma=0,1$, is a specified subset of nodes of $G$.  For
the $D_4$ example discussed below, $b=1$ and $T_\gamma$ is the set of
nodes of ${\Bbb Z}_2$ grading equal to $\gamma$ so $T_0=\{2\}$ and
$T_1=\{1,3,4\}$.  For the $A$-$D$-$E$\, WZW theories
$n_{ia}{}^b=F^i_{ab}$, however, since the minimal models are WZW
cosets there is an additional tensor product structure of the graphs
$A\otimes G$ giving~\cite{PZ0011021}
\begin{equation}
       {\tilde V}_{(r,s)\,(r',s');1}{}^{(r'',a,b,\zeta)} =
N^{(A_{g-2})}_{rr'}{}^{r''}\, {\tilde V}_{ij^*;1}^{(G)}{}^x
\label{minVtilde}
\end{equation}
where $N_r=F^r$ are the $A$-type Verlinde fusion matrices of
$\hat{sl}(2)_{g-2}$.  The integers ${\tilde
V}^{(G)}_{ij^*;1}{}^x\in{\Bbb N}$ encode~\cite{Ocn,BE,BEK99,BEK00} the
Ocneanu quantum graphs and the fusion algebra of the WZW models with a
seam.  For the minimal models the fusion algebra and quantum graphs
are encoded by (\ref{minVtilde}).

For diagonal $A$-type theories the conformal twisted partition
functions simplify to
\begin{equation}
Z_k(q)=\sum_{i,j} N_{ij}{}^k
\chi_{i}(q){\chi_j(q)}^*
\end{equation}
In particular, for $(A_{L\mi1},A_{L})$ unitary minimal models the
seams are labelled by the Kac labels $(r,s)$ and, since the ${\Bbb
Z}_2$ diagram automorphism is included in the fusion algebra, we can
take $\zeta=1$.  In this case the partition functions are given in
terms of Virasoro characters by~\footnote{This formula for the case
$r=r'=r''=1$ was obtained by Orrick and Pearce and explained in
private communication to Zuber in September 1999 at the time of the
$A$-$D$-$E$ Conference in Warwick.}
\begin{equation}
Z_{(r,s)}(q)=\sum_{(r',s'),(r'',s'')} N^{(A_{g-2})}_{r'r''}{}^r 
N^{(A_{g-1})}_{s's''}{}^s
\chi_{r',s'}(q){\chi_{r'',s''}(q)}^*
\end{equation}
These results were verified numerically for $L=3, 4, 5$ and $6$ and 
matrix size given by $N=22,\,16,14$ and $12$ respectively.

The integrable seam weights of the lattice models give the physical
content of the conformal twisted boundary conditions.  This physical
content is not at all clear from the conformal labels alone.  We
illustrate this by discussing the Ising and 3-state Potts models as
examples.
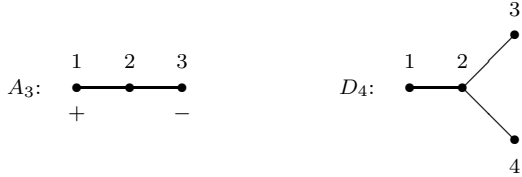
\begin{center}
\newcommand{\spos}[2]{\makebox(0,0)[#1]{{$\scriptstyle #2$}}}
\newcommand{\sm}[1]{{\scriptstyle #1}}
\setlength{\unitlength}{7mm}
     \begin{picture}(4.5,2.3)(0,-.5)
	\put(-1,0){\spos{}{A_{3}:}}
\put(0,0){\line(1,0){2}}
\multiput(0,0)(1,0){3}{\spos{}{\bullet}}
\put(0,0.5){\spos{}{1}}
\put(0,-0.5){\spos{}{+}}
\put(1,0.5){\spos{}{2}}
\put(2,0.5){\spos{}{3}}
\put(2,-0.5){\spos{}{-}}
\end{picture}
\hspace{1cm}
\begin{picture}(4.5,2.3)(0,-.5)
	\put(-1,0){\spos{}{D_{4}:}}
\put(0,0){\line(1,0){1}}
\multiput(0,0)(1,0){2}{\spos{}{\bullet}}
\put(1,0){\line(1,1){1}}
\put(1,0){\line(1,-1){1}}
\put(2,1){\spos{}{\bullet}}
\put(2,-1){\spos{}{\bullet}}
\put(0,0.5){\spos{}{1}}
\put(1,0.5){\spos{}{2}}
\put(2,1.5){\spos{}{3}}
\put(2,-1.5){\spos{}{4}}
\end{picture}
\end{center}

\subsection{Ising $(A_2,A_3)$ example}

The three twisted partition functions $Z_{(r,s)}$ for the Ising model are
\begin{eqnarray}
Z_{P}=Z_{0}\,\,=Z_{(1,1)}
&=&|\chi_{0}(q)|^{2}+|\chi_{\frac{1}{2}}(q)|^{2}+|\chi_{\frac{1}{16}}(q)|^{2
       }\\[2pt]
Z_{A}=Z_{\frac{1}{2}}\,=Z_{(1,3)}
       &=&\chi_{0}(q){\chi_{\frac{1}{2}}(q)}^{*}
       +{\chi_{0}(q)}^{*}\chi_{\frac{1}{2}}(q)+|\chi_{\frac{1}{16}}(q)|^{2}
       \\[2pt]
Z_{\frac{1}{16}}=Z_{(1,2)}
       &=&[\chi_{0}(q)+\chi_{\frac{1}{2}}(q)]{\chi_{\frac{1}{16}}(q)}^{*}+
       {[\chi_{0}(q)+\chi_{\frac{1}{2}}(q)]^{*}}\chi_{\frac{1}{16}}(q)
\label{IsingPFs}
\end{eqnarray}
The $(r,s)=(1,1)$ and $(1,3)$ seams reproduce the Periodic $P$ and
Antiperiodic $A$ boundary conditions respectively and their associated
partition functions~\cite{Car84}.  The weights giving the physical
content of the third seam are
\begin{eqnarray}
  \begin{array}{rclcrcl}
            W^{(1,2)} \W{\pm \\&\pm}{.} & = & i e^{i 
\frac{\pi}{8}}-\sqrt{2}\,i e^{-i \frac{\pi}{8}},
&\qquad\quad  &
	    W^{(1,2)}\W{\pm \\&\mp}{.} & = & i e^{i \frac{\pi}{8}}\\[14pt]
       W^{(1,2)}\W{&\pm\\ \pm}{.} & = & i e^{i \frac{\pi}{8}}- i e^{-i 
\frac{\pi}{8}}/\sqrt{2}, &  &
	    W^{(1,2)}\W{&\pm\\ \mp}{.} & = & -i e^{-i \frac{\pi}{8}}/\sqrt{2}
	\end{array}
         \label{eq:Ising12}
\end{eqnarray}
where the upper or lower signs are taken and for $A_3$ we identify
$+=1$, $-=3$ and the frozen state $0=2$ is omitted.  The $(1,2)$ seam
weights are complex but at the isotropic point $u=\pi/8$ the $(1,2)$
transfer matrix is hermitian so its eigenvalues are real.  In fact,
after removing $\pm$ degeneracies, the eigenvalues are all positive.

The twisted Ising partition functions are obtained numerically to very
high precision.  The energy levels in the $q$ series of the twisted
partition functions are reproduced for at least the first 10 levels
counting degeneracies to 4--8 digit accuracy.  The conformal weights
are obtained to 8 digit accuracy.

\subsection{3-state Potts $(A_4,D_4)$ example}

The conformal twisted partition functions of the $3$-state Potts model
have been listed by Petkova and Zuber~\cite{PZ0011021}.  This list
extends the previously known twisted boundary
conditions~\cite{vGR,Car86,Zub86} corresponding to the automorphisms
$\zeta=1,\omega,\tau\in\Gamma(D_4)$ and corresponding to the periodic
$P=(1,1,1)$, cyclic $C=(1,1,\omega)$ and twisted $T=(1,1,\tau)$
boundary conditions respectively.  Explicitly, the 3- and 2-cycles are
given by the permutations $\omega=(1,3,4)$ and $\tau=(3,4)$.

The twisted partition functions are written most compactly in terms of
the extended block characters
\begin{equation}
\hat\chi_{r,a}(q)=\sum_{s\in A_{g-1}} n_{s1}{}^a\chi_{r,s}(q)
=\sum_{s\in A_{g-1}} F_{1a}^s\chi_{r,s}(q)
\end{equation}
where $\chi_{r,s}(q)$ are the Virasoro characters.  Considering all
the seams $(r,s,\zeta)$ and taking into account symmetries, we find 8
distinct conformal twisted partition functions in complete agreement
with Petkova and Zuber~\cite{PZ0011021}
\begin{eqnarray}
Z_P=Z_{(1,1,1)}
 &\!\!\!=&|\hat\chi_{1,1}(q)|^{2}+|\hat\chi_{1,3}(q)|^{2}+|\hat\chi_{1, 
4}(q)|^{2}
 +|\hat\chi_{3,1}(q)|^{2}+|\hat\chi_{3,3}(q)|^{2}
 +|\hat\chi_{3,4}(q)|^{2}\notag\\[6pt]
Z_{(1,2,1)}
       &\!\!\!=&\hat\chi_{1,2}(q)
 {[\hat\chi_{1,1}(q)+\hat\chi_{1,3}(q)+\hat\chi_{1,4}(q)]^{*}}
 \notag\\[1pt]
&&\phantom{=}\;
+\hat\chi_{3,2}(q)
 {[\hat\chi_{3,1}(q)+\hat\chi_{3,3}(q)+\hat\chi_{3,4}(q)]^{*}}
 \notag\\[6pt]
Z_C=Z_{(1,1,\omega)}
       &\!\!\!=&\hat\chi_{1,1}(q)\hat\chi_{1,3}(q)^{*}
 +\hat\chi_{1,1}(q)^{*}\hat\chi_{1,3}(q)+|\hat\chi_{1,3}(q)|^{2}
 \notag\\[1pt]
&&\phantom{=}\;
+\hat\chi_{3,1}(q)\hat\chi_{3,3}(q)^{*}
 +\hat\chi_{3,1}(q)^{*}\hat\chi_{3,3}(q)+|\hat\chi_{3,3}(q)|^{2}
 \notag\\[6pt]
Z_T=Z_{(1,1,\tau)}
       &\!\!\!=&|\hat\chi_{1,2}(q)|^{2}+|\hat\chi_{3,2}(q)|^{2}
       \notag\\[6pt]
Z_{(3,1,1)}
       &\!\!\!=&\hat\chi_{1,1}(q)\hat\chi_{3,1}(q)^{*}
       +\hat\chi_{1,1}(q)^{*}\hat\chi_{3,1}(q)+|\hat\chi_{3,1}(q)|^{2}
       \\[1pt]
&&\phantom{=}\;
+\hat\chi_{1,3}(q)\hat\chi_{3,3}(q)^{*}
 +\hat\chi_{1,3}(q)^{*}\hat\chi_{3,3}(q)+|\hat\chi_{3,3}(q)|^{2}
 \notag\\[1pt]
&&\phantom{=}\;
+\hat\chi_{1,4}(q)\hat\chi_{3,4}(q)^{*}
 +\hat\chi_{1,4}(q)^{*}\hat\chi_{3,4}(q)+|\hat\chi_{3,4}(q)|^{2}
 \notag
     \end{eqnarray}
\begin{eqnarray}
Z_{(3,2,1)}
&\!\!\!=&\hat\chi_{1,2}(q)[\hat\chi_{3,1}(q)+\hat\chi_{3,3}(q)+
\hat\chi_{3,4}(q)]^{*}\notag\\[1pt]
&&\phantom{=}\;
+\hat\chi_{3,2}(q)[\hat\chi_{1,1}(q)+\hat\chi_{1,3}(q)
+\hat\chi_{1,4}(q)
       +\hat\chi_{3,1}(q)+\hat\chi_{3,3}(q)+\hat\chi_{3,4}(q)]^{*}
       \notag\\[6pt]
Z_{(3,1,\omega)}
&\!\!\!=&\hat\chi_{3,3}(q)[\hat\chi_{1,1}(q)+\hat\chi_{3,1}(q)+
\hat\chi_{1,4}(q)]^{*}
+\hat\chi_{1,3}(q)\hat\chi_{3,1}(q)^{*}\notag\\[1pt]
&&\phantom{=}\;
+\hat\chi_{3,3}(q)^{*}[\hat\chi_{1,1}(q)+\hat\chi_{3,1}(q)
+\hat\chi_{1,4}(q)]
+\hat\chi_{1,3}(q)^{*}\hat\chi_{3,3}(q)+|\hat\chi_{3,3}(q)|^{2}
\notag\\[6pt]
Z_{(3,1,1)}
&\!\!\!=&\hat\chi_{1,2}(q)\hat\chi_{3,2}(q)^{*}
+\hat\chi_{3,2}(q)\hat\chi_{1,2}(q)^{*}
+|\hat\chi_{3,2}(q)|^{2}\notag
     \end{eqnarray}
Our construction labels $(r,s,\zeta)$ correspond with the labels
$(r,a,\gamma)$ of Petkova and Zuber with the obvious identifications
$\zeta=1\mapsto\gamma=1$, $\zeta=\tau\mapsto\gamma=0$ and
$(r,1,\omega)\mapsto (r,3,1)$.

The twisted 3-state Potts partition functions are obtained numerically
with reasonable precision and, as expected, intertwine with those of
$A_5$.  The energy levels in the $q$ series of the twisted partition
functions are reproduced for at least the first 10 levels counting
degeneracies to 2--4 digit accuracy.  The conformal weights are
obtained to at least 4 digit accuracy.

The other $D$ and $E$ cases are of much interest because of
connections with Ocneanu quantum graphs.  We will present the details
of these cases in our subsequent paper~\cite{CMOP2}.  Although we
have emphasized the specialized conformal twisted boundary conditions
in this letter, we point out that the same fusion techniques can be
used to construct integrable seams off-criticality for the elliptic
$A$ and $D$ lattice models.

\section*{Acknowledgements} 			\label{sec:Acknowldegments}
This research is supported by the Australian Research Council.  We
thank Jean-Bernard Zuber for continued interest, careful reading and
encouragement.

\bibliographystyle{UNSRT} \bibliography{seam}

\end{document}